\documentclass[english]{article}
\usepackage[T1]{fontenc}
\usepackage[utf8]{inputenc}
\usepackage[a4paper]{geometry}
\usepackage{amsmath}
\usepackage{amssymb}
\usepackage{graphicx}
\usepackage{esint}
\usepackage{xspace}
\usepackage{lineno}
\usepackage{setspace}
\usepackage{url}
\usepackage{color}
\usepackage{hyperref}
\usepackage[numbers,sort&compress]{natbib}

\makeatletter

\newcommand{\CCFM}{Ciafaloni:1987ur,Catani:1989yc,Catani:1989sg,CCFMd}

\usepackage[normalem]{ulem}\usepackage{color}
\usepackage{hyperref}

\usepackage[numbers,sort&compress]{natbib}

\newcommand{\chisq}{\ensuremath{\chi^{2}\!/\mathrm{NDP}}\xspace}

\makeatother

\usepackage{babel}

\graphicspath{{figs/}}

\begin{document}

\title{Unintegrated gluon distributions for forward jets at LHC}

\date{}
\author{Piotr Kotko$^1$, Wojciech S\l{}omi\'{n}ski$^2$ and Dawid Toton$^3$ \\\\
$^1$ {\small\it Department of Physics, Penn State University,}\\ {\small\it University Park, 16803 PA, USA}\\\\
$^2$ {\small\it The M. Smoluchowski Institute of Physics, Jagiellonian University,}\\{\small\it  S. Łojasiewicza 11, 30-348 Kraków, Poland}\\\\
$^3$ {\small\it The H.\ Niewodnicza\'nski Institute of Nuclear Physics, Polish Academy of Sciences,}\\{\small\it Radzikowskiego 152, 31-342 Krak\'ow, Poland}
}

\maketitle

\begin{abstract}
We test several BFKL-like evolution equations for unintegrated gluon
distributions against forward-central dijet production at LHC. Our study
is based on fitting the evolution scenarios to the LHC data
using the high energy factorization approach.
Thus,
as a by-product, we obtain a set of 
LHC-motivated
 unintegrated gluon distributions ready
to use. We utilize this application by calculating azimuthal decorrelations
for forward-central dijet production and compare with existing data. 
\end{abstract}

\section{Introduction}

A typical procedure in applying QCD to hadronic collisions relies
on factorization theorems. They consist in two ingredients: a perturbatively
calculable hard part and a nonperturbative piece parametrizing hadrons
participating in a collision. The most known and tested is the collinear
factorization (see e.g. \citep{Collins:2011zzd} for a review), which applies for a
variety of processes, including jet observables in deep inelastic
scattering (DIS) and hadron-hadron collisions. Here, the nonperturbative
component is parametrized in terms of parton distribution functions
(PDFs) which undergo Dokshitzer-Gribov-Lipatov-Altarelli-Parisi (DGLAP)
evolution equations. The key feature of PDFs is the \textit{universality},
i.e. the PDFs that are measured in one process can be used in any
other for which the factorization holds. Therefore, for instance one
can use PDFs fitted to DIS structure functions and use them to make
predictions for jets in hadron-hadron collisions. Although the collinear
factorization is powerful and well-tested, it is supposed that for
certain observables, e.g. forward jets at high energies, another kind
of evolution equations for the PDFs is needed. 
 Namely, the perturbative calculations contain 
the logarithms of the form $\alpha_{s}\log(1/x)$, where
 $x$ is the longitudinal fraction
of the hadron momentum carried by the parton. At high energies and forward rapidities
$x$ is small and these logarithms need to be resumed.
This is accomplished by means of various
``small~$x$'' evolution equations, which essentially are various extensions of the pioneering 
Balitski-Fadin-Kuraev-Lipatov (BFKL) evolution equation
(see e.g. \cite{Lipatov:1996ts}).
In the small $x$ domain the transverse momenta of the partons exchanged between the perturbative and nonperturbative parts are not suppressed comparing to the collinear factorization. 
Therefore, the PDFs have an explicit dependence on the transverse momentum of a parton. Such objects are often referred to, as
transverse momentum dependent PDFs (TMDs) or Unintegrated
PDFs, although the former are typically used outside the small $x$ physics, and posses unambiguous (though in general process dependent) field theoretic definitions.
Actually, at small $x$ one usually deals with initial state
gluons only, and thus the object of interest in this paper is an Unintegrated
Gluon Distribution (UGD). The UGDs have to
be convoluted with a perturbative ``hard part'' according to so-called
$k_{T}$ or High Energy Factorization (HEF). We describe this approach
in some more detail in Section~\ref{sec:HEF}. Here, let us just
mention that unlike the collinear factorization, the HEF is not a
QCD theorem and actually the universality of UGDs is supposed to be
violated for jet production in hadron-hadron collisions. 
Thus, in principle, the standard procedure of fitting the
UGDs to the $F_{2}$ HERA
data and using it for jets in hadron-hadron
collisions is not correct, but there are no quantitative measures
of the factorization violation so far. 
Actually, HEF is surprisingly quite successful with describing LHC data using UGDs from fits to structure functions, see for instance~\citep{vanHameren:2014ala}.
At present, there are several fits to $F_{2}$
data using different small~$x$ approaches, see \citep{Ellis:2008yp,Ross:2011zzb,Kutak:2012rf,Lipatov:2013yra}
for more details.

In the present work we undertake another path. We make an attempt to fit various BFKL-like UGDs directly
to the LHC data for jet forward jet production. It has a twofold purpose. First, we have an opportunity to explore
UGDs using relatively exclusive observables. Second, we want to free ourselves from the aforementioned universality problem when transferring UGDs from DIS to the LHC domain. 
We consider two separate measurements: jet transverse momentum spectra~\citep{Chatrchyan2012} 
in forward-central jet production and forward-central dijet decorrelations~\citep{CMS:2014oma}.
The first measurement consists of two separate sets of data: for the forward jet and for the
central jet. Thus, the mutual description of both spectra imposes
a strong constraint on the UGDs and we shall use this measurements to make our fits.
The second measurement will be used to test the fits.

The paper is organized as follows. In Section~\ref{sec:HEF} we describe the approach of HEF. The small $x$ evolution equations with various components incorporating sub-leading effects are discussed in Section~\ref{sec:evolution_eqs}.
The fitting procedure and the software used
are described in Section~\ref{sec:Procedure}. We give the results
in Section~\ref{sec:Results}. Having the fits, we test them against
recent forward-central dijet decorrelations data
in Section~\ref{sec:decorrelations}. Finally, we discuss our research in Section~\ref{sec:Summary}.

\section{High Energy Factorization}

\label{sec:HEF}

In this introductory section we discuss in more detail issues
concerning factorization at small $x$. This task is somewhat complicated,
notably because of the various existing approaches and various existing
definitions of UGDs. 

In the following paper the notion of HEF corresponds to a general class
of factorization approaches supposed to be valid at small $x$. Below
we list some of the existing realizations:
\begin{enumerate}
\item the factorization of Gribov, Levin and Ryskin (GLR) \citep{Gribov1983}
for high-$p_{T}$ inclusive gluon production
\label{HEFGLR}
\item the factorization of Catani, Ciafaloni and Hautmann (CCH) \citep{Catani:1990eg,Catani:1994sq}
for heavy quark production in DIS, photo-production and hadron-hadron
collisions
\label{HEFCCH}
\item the factorization of Collins and Ellis \citep{Collins1991} for
heavy quark production in hadron-hadron collisions
\label{HEFCE}
\item the factorization for inclusive gluon production in the saturation
regime for proton-nuclei collisions within the Color Glass Condensate (CGC)
approach \citep{Blaizot:2004wu} and color dipole formalism \citep{Kovchegov:2001sc,Nikolaev:2004cu} (the equivalence of both approaches
was shown in \citep{Iancu2004}) 
\label{HEFCGC}
\end{enumerate}
In these approaches the nonperturbative part is parametrized in terms of UGDs undergoing BFKL evolution (for GRL, CCH, Collins-Ellis factorizations) or nonlinear Balitsky-Kovchegov evolution \citep{Balitsky:1995ub,Kovchegov:1999yj} (for CGC).  On the other
hand, superficially similar objects to UGDs appear in so-called transverse
momentum dependent (TMD) factorization and are called TMD PDFs. One
should however realize that the enumerated approaches are valid at leading logarithmic approximation, while the
TMD factorizations are valid to all orders in the leading twist approximation. 
Moreover, unlike most of UGDs in the HEF factorizations,
the TMD PDFs have precise operator definitions in terms of matrix elements
of nonlocal operators. Those definitions require appropriate Wilson
lines to be inserted in order to make the definitions gauge invariant
and to resum collinear gluons related to final and initial state interactions.
These insertions make the TMD PDFs, in general, process dependent and
thus non-universal, breaking the principle of factorization (for more
details see e.g. \citep{Bomhof:2006dp,Mulders:2011zt}). Only for
processes with at most two hadrons the TMD factorization is proved
to hold to all orders (for example back-to-back single hadron production in DIS or Drell-Yan
scattering). The natural question arises whether the non-universality
of TMD PDFs transfers to the small $x$ limit. In ref. \citep{Xiao:2010sp}
an explicit arguments were given that this is the case for dilute-dense collisions
(actually the arguments hold for so-called ``hybrid'' factorization
-- see also below). Moreover it is known from the CGC approach that at really small $x$, i.e. 
in the saturation regime, the cross sections cannot be described by
just dipoles (averages of two Wilson lines), but also higher correlators
are needed \citep{Dominguez:2012ad}, what violates the ordinary logic of factorization. However,
for the case of back-to-back dijet production in dilute-dense collisions a
generalized factorization has been proposed \citep{Dominguez:2011wm};
that is, the cross section can be given in terms of hard factors and
certain universal pieces. Recently, these results were improved to the case
of imbalanced dijets \citep{Kotko:2015ura}. In particular, when the imbalanced transverse momentum is of the order 
of transverse momenta of the jets the HEF for dijet production can be derived from the dilute limit of the CGC approach.

In the present work we shall constrain ourselves to dijet production
in p-p collisions in the linear regime, as the kinematics we are interested
in (and where the data exist) do not allow to develop the saturation
region. We want to utilize most of the phase space covered by the
data, thus we do not constrain ourselves to the back-to-back dijet
region analyzed in \citep{Dominguez:2011wm}. Rather, we shall use the HEF factorization for dijet production. Since this approach is an extension of the CCH formalism, we shall now briefly recall the latter and the required extensions to obtain HEF for dijets. For a direct derivation from CGC approach see \citep{Kotko:2015ura}.

\begin{figure}
\begin{center}
\parbox{5cm}{A)\\\includegraphics[width=5cm]{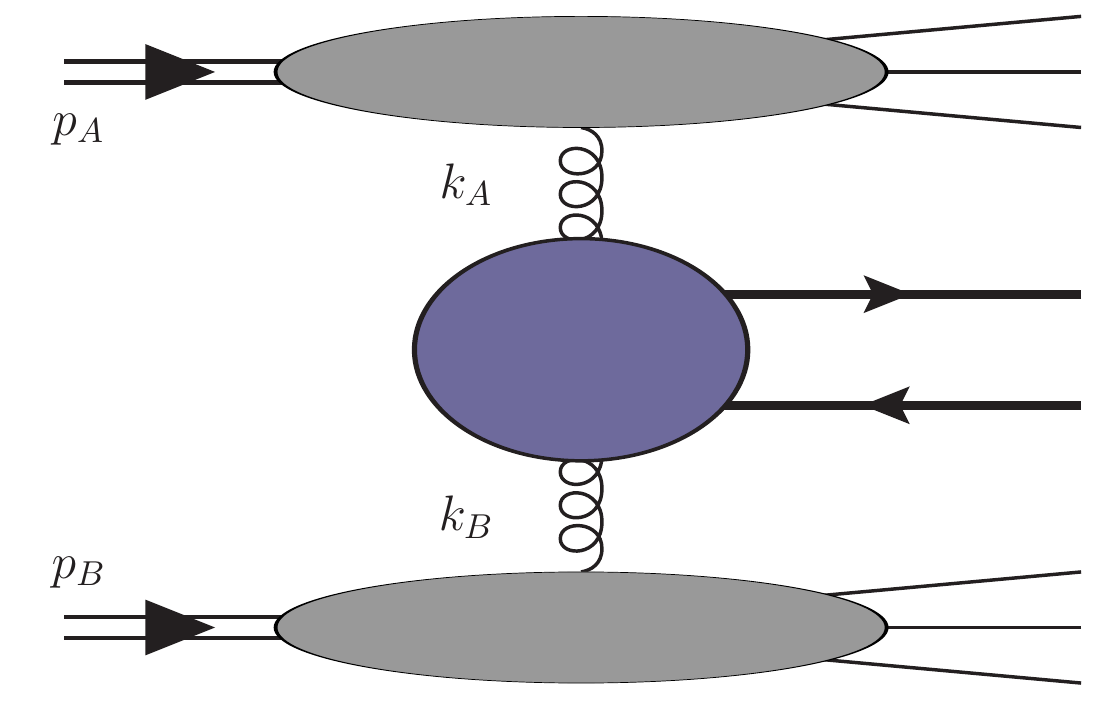}}~~~~~~~~~~~~\parbox{5cm}{B)\\\includegraphics[width=5cm]{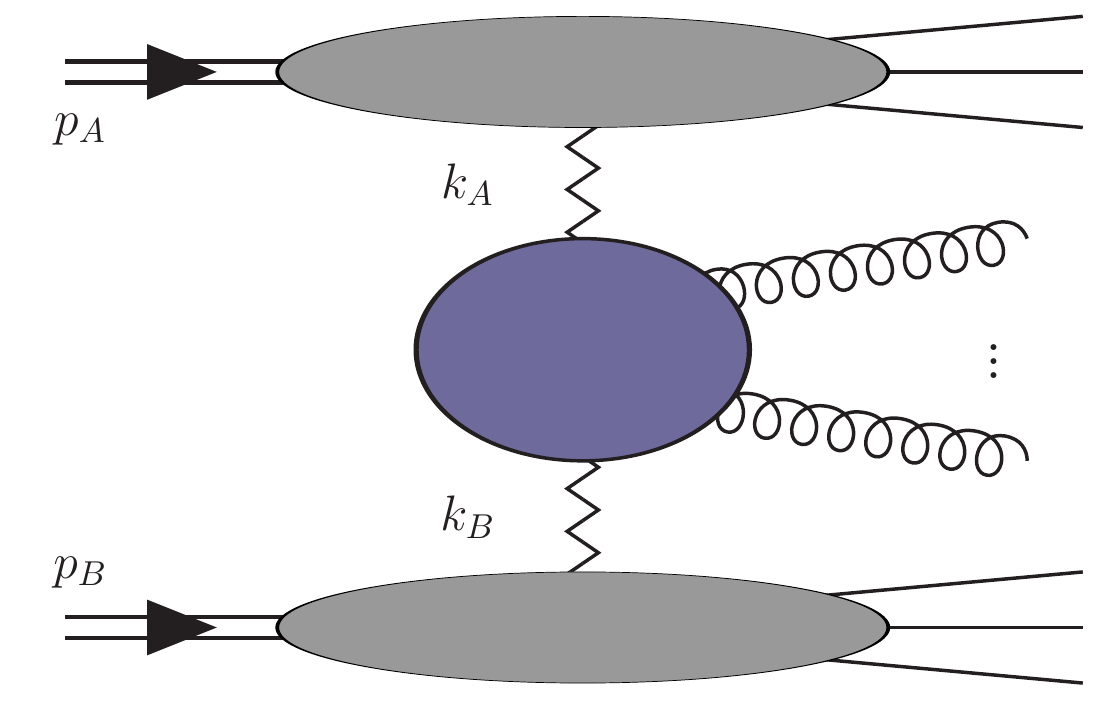}}
\par\end{center}

\caption{A) The CCH factorization for inclusive heavy quark production; despite
the fact that the gluons entering the central blob are off-shell the
sub-process is gauge invariant. B) For sub-processes with final state
gluons the gauge invariance requires the off-shell gluons to be replaced
by the effective particles giving rise to multiple eikonal gluon exchanges
between the blobs. \label{fig:HEF}}
\end{figure}

In the CCH high energy factorization, one considers the heavy quark
pair produced via the tree-level hard sub-process $g^{*}\left(k_{A}\right)g^{*}\left(k_{B}\right)\rightarrow Q\overline{Q}$
in the axial gauge. The initial state gluons are off-shell and have
the momenta of the form $k_{A}=x_A\, p_{A}+k_{TA}$ and $k_{B}=x_B\, p_{B}+k_{TB}$,
where $p_{A}$, $p_{B}$ are the momenta of the incoming hadrons and
$p_{A}\cdot k_{TA}=p_{B}\cdot k_{TB}=0$. This particular form of
the exchanged momenta is a result of the imposed high energy limit.
The off-shell gluons have ``polarization vectors'' that are $p_{A}$
and $p_{B}$ respectively. Thanks to this kinematics the sub-process
given by ordinary Feynman diagrams is gauge invariant despite its
off-shellness. In CCH approach the factorization formula for heavy
quark production reads (see Fig. \ref{fig:HEF}A)
\begin{multline}
d\sigma_{AB\rightarrow Q\overline{Q}}=\int d^{2}k_{T A}\int\frac{dx_{A}}{x_{A}}\,\int d^{2}k_{T B}\int\frac{dx_{B}}{x_{B}}\,\\
\mathcal{F}_{g^{*}/A}\left(x_{A},k_{T A}\right)\,\mathcal{F}_{g^{*}/B}\left(x_{B},k_{T B}\right)\, d\hat{\sigma}_{g^{*}g^{*}\rightarrow Q\overline{Q}}\left(x_{A},x_{B},k_{T A},k_{T B}\right),\label{eq:HEN_fact_1}
\end{multline}
where $d\hat{\sigma}_{g^{*}g^{*}\rightarrow Q\overline{Q}}$ is the
partonic cross section build up from the gauge invariant $g^{*}g^{*}\rightarrow Q\overline{Q}$
amplitude and $\mathcal{F}_{g^{*}/A}$, $\mathcal{F}_{g^{*}/B}$ are
UGDs for hadrons $A$ and $B$. The contributions with off-shell quarks
are suppressed. The UGDs are assumed to undergo the BFKL evolution
equations. In Ref. \citep{Catani:1994sq} it was argued that similar
factorization holds to all orders for DIS heavy quark structure function,
although the argumentation misses the details comparing to collinear
factorization proofs~\citep{Collins:2011zzd}, especially the definitions
of UGDs and complications arising at higher orders in the axial gauge~\citep{Avsar:2012hj}.

In the works \citep{Deak2010},\citep{Kutak:2012rf},\citep{VanHameren2013},\citep{vanHameren2013a,vanHameren:2014lna,vanHameren:2014ala}
as well as in this paper the CCH factorization was extended to model
the cross section for jet production in hadron-hadron collisions.
The first difficulty arises because now one has to consider also gluons
in the final state, e.g. $g^{*}g^{*}\rightarrow gg$ sub-process for
dijet production. The corresponding amplitude is however not gauge invariant
when calculated from ordinary Feynman diagrams. A few approaches have
been proposed to calculate a gauge invariant extension of such amplitudes \citep{vanHameren2012,vanHameren2013a,vanHameren:2013csa,vanHameren:2014iua,Kotko2014a}.
These {\it gauge invariant off-shell amplitudes} in fact correspond to a vertex that can be calculated from
the well-known Lipatov's effective action \citep{Lipatov:1995pn,Antonov:2004hh}
(see Fig. \ref{fig:HEF}B). The approaches \citep{vanHameren2012,vanHameren2013a,vanHameren:2013csa,vanHameren:2014iua,Kotko2014a} were however oriented on practical and efficient computations of multi-particle off-shell amplitudes
using helicity method and computer codes. As stated before, in CCH the UGDs undergo BFKL
evolution. In our extensions of CCH approach we allow the UGDs to undergo
more complicated evolution equations, which are more suitable for
jets. More details will be given in Section \ref{sec:evolution_eqs}.
Yet another modification of the CCH formula comes from the fact that
the present study concerns the system of dijets where one of the jet
is forward, while the second is in the central region. From $2\rightarrow2$
kinematics it follows then, that $x_{A}\ll x_{B}$ (or the opposite),
except for the small corner of the phase space. Since $x_{B}$ is
typically of the order of $0.5$ the usage of small $x$ evolution
for $\mathcal{F}_{g^{*}/B}$ is questionable (this is similar to dilute-dense
system considered e.g. in \citep{Dominguez:2011wm}). Therefore we
use collinear approach on the $B$ hadron side \citep{Deak:2009xt}.
Technically, one takes the collinear limit in $d\hat{\sigma}_{g^{*}g^{*}\rightarrow2j}$
by sending $k_{TB}\rightarrow0$ to obtain a sub-process with one off-shell
gluon $d\hat{\sigma}_{g^{*}g\rightarrow2j}$ (the off-shell amplitudes
have well defined on-shell limit). In this limit one has 
to take into account also sub-processes with initial state on-shell
quarks, $d\hat{\sigma}_{g^{*}q\rightarrow2j}$. The remaining integral
over $d^{2}k_{TB}$ gives helicity sum for $B$ partons on one hand,
and the integrated (collinear) PDF on the other $\int dk_{B}^{2}\,\mathcal{F}_{a^{*}/B}\left(x_{B},k_{TB}\right)=f_{a}\left(x_{B}\right)$.
Thus, the final formula for the factorization model reads
\begin{multline}
d\sigma_{AB\rightarrow2j}=\int d^{2}k_{TA}\int\frac{dx_{A}}{x_{A}}\,\int\frac{dx_{B}}{x_{B}}\,\sum_{b}\\
\mathcal{F}_{g^{*}/A}\left(x_{A},k_{T A},\mu\right)\, f_{b}\left(x_{B},\mu\right)\, d\hat{\sigma}_{g^{*}b\rightarrow2j}\left(x_{A},x_{B},k_{T A},\mu\right),\label{eq:HEN_fact_2}
\end{multline}
where we have included the hard scale dependence not only in the collinear
PDFs $f_{b}$, but in the UGD as well. Such a dependence turns out
to be important for certain exclusive observables involving a hard
scale (e.g. large $p_{T}$ of jets; see e.g. \citep{vanHameren:2014ala}).
We note, that when the final states become well separated in rapidity,
i.e. when the central jet lies in the opposite hemisphere to the forward
jet we start to violate our condition $x_{A}\ll x_{B}$ and different
approach should be used. The factorization formula (\ref{eq:HEN_fact_2})
resembles the linearized approach of \citep{Dominguez:2011wm} but
it extends beyond the correlation limit as here the hard sub-processes
have injected a nonzero $k_{T}$. As mentioned before, the formula (\ref{eq:HEN_fact_2})
has been recently derived from the CGC approach in \citep{Kotko:2015ura}.

\section{Small \texorpdfstring{$x$}{\textit{x}} evolution equations }

\label{sec:evolution_eqs}

Let us now discuss the evolution equations for UGDs which were used
in our fits. As described in the preceding section we concentrate
on linear evolution equations. Below we list some of them with a short
explanation. We consider only gluon UGDs, thus we skip the subscripts
in $\mathcal{F}_{g^{*}/A}$.
\begin{enumerate}
\item pure BFKL equation\label{enu:pure-BFKL-equation}

The equation in the leading logarithmic approximation reads \citep{Fadin:1975cb,Kuraev:1977fs}
\begin{multline}
\mathcal{F}\left(x,k_{T}^{2}\right)=\mathcal{F}_{0}\left(x,k_{T}^{2}\right)\\
+\overline{\alpha}_{s}\int_{x}^{1}\frac{dz}{z}\,\int_{0}^{\infty}dq_{T}^{2}\left[\frac{q_{T}^{2}\mathcal{F}\left(\frac{x}{z},q_{T}^{2}\right)-k_{T}^{2}\mathcal{F}\left(\frac{x}{z},k_{T}^{2}\right)}{\left|q_{T}^{2}-k_{T}^{2}\right|}+\frac{k_{T}^{2}\mathcal{F}\left(\frac{x}{z},k_{T}^{2}\right)}{\sqrt{4q_{T}^{4}+k_{T}^{4}}}\right]\label{eq:BFKL}
\end{multline}
where $\overline{\alpha}_{s}=N_{c}\alpha_{s}/\pi$ with $N_{c}$ being
the number of colors. The initial condition for the evolution is given
by $\mathcal{F}_{0}$. The NLO BFKL equation is also known \citep{Fadin:1998py,Ciafaloni:1998gs}.
One of the drawbacks of the pure BFKL equation comes from the fact that
$q_{T}^{2}$ of the gluons emitted along the ladder is unconstrained.
Indeed, since in the BFKL regime the virtuality of the exchanged gluons
is dominated by the transverse components, the resulting \textit{kinematic
constraint} reads \citep{Andersson:1995jt,Kwiecinski:1996a}
\begin{equation}
q_{T}^{2}<\frac{1-z}{z}\, k_{T}^{2}\approx\frac{1}{z}\, k_{T}^{2}.\label{eq:kinematic_constr}
\end{equation}
This constraint is also often referred to as the consistency constraint.

\item BFKL with the kinematic constraint (BFKL+C)

To incorporate the consistency constraint one may include the appropriate
step function into the real emission part of the BFKL. This operation,
actually introduces some higher order corrections into the BFKL equation
\citep{Kwiecinski:1996a}. In addition, one may introduce another
class of sub-leading corrections by allowing the strong coupling constant
to run with the local scale along the ladder. Finally, one may define
the $q_{T}^{2}$ integration region to lie away from the infrared
nonperturbative region by separating the $\int_{0}^{k_{T0}^{2}}dq_{T}^{2}$
integration and moving it to the initial condition (the infrared cutoff
$k_{T0}^{2}$ is taken to be of the order of $1\,\mathrm{GeV}$).
The improved equation reads \citep{Kwiecinski:1997ee} 
\begin{multline}
\mathcal{F}\left(x,k_{T}^{2}\right)=\mathcal{F}_{0}\left(x,k_{T}^{2}\right)\\
+\overline{\alpha}_{s}\left(k_{T}^{2}\right)\int_{x}^{1}\frac{dz}{z}\,\int_{k_{T0}^{2}}^{\infty}dq_{T}^{2}\left[\frac{q_{T}^{2}\mathcal{F}\left(\frac{x}{z},q_{T}^{2}\right)\Theta\left(k_{T}^{2}-zq_{T}^{2}\right)-k_{T}^{2}\mathcal{F}\left(\frac{x}{z},q_{T}^{2}\right)}{\left|q_{T}^{2}-k_{T}^{2}\right|}+\frac{k_{T}^{2}\mathcal{F}\left(\frac{x}{z},k_{T}^{2}\right)}{\sqrt{4q_{T}^{4}+k_{T}^{4}}}\right].\label{eq:BFKL_constr}
\end{multline}
Recently, it has been studied in the context of the Mueller-Navelet jets,
that the energy-momentum conservation violation (which above is cured
by a ``brute force'') becomes less harmful when full NLO corrections
are applied \citep{Ducloue:2014koa}. The effects of the kinematic constraints in 
the approximate form (\ref{eq:kinematic_constr}) as well as in the full form have been recently 
analyzed \citep{Deak:2015dpa} in the context of the CCFM evolution equation  \cite\CCFM. 

\item BFKL with the kinematic constraint in re-summed form (BFKL+CR)

The equation (\ref{eq:BFKL_constr}) can be casted in yet another
form \citep{Kutak:2011fu}
\begin{multline}
\mathcal{F}\left(x,k_{T}^{2}\right)=\tilde{\mathcal{F}}_{0}\left(x,k_{T}^{2}\right)\\
+\overline{\alpha}_{s}\left(k_{T}^{2}\right)\int_{x}^{1}\frac{dz}{z}\,\int_{k_{T0}^{2}}^{\infty}\frac{d^{2}q_{T}}{\pi q_{T}^{2}}\Theta\left(q_{T}^{2}-\mu^{2}\right)\Delta_{R}\left(z,k_{T}^{2},\mu^{2}\right)\mathcal{F}\left(\frac{x}{z},\left|\vec{k}_{T}+\vec{q}_{T}\right|^{2}\right),\label{eq:BFKL+CR}
\end{multline}
where 
\begin{equation}
\Delta_{R}\left(z,k_{T}^{2},\mu^{2}\right)=\exp\left(-\overline{\alpha}_{s}\ln\frac{1}{z}\,\ln\frac{k_{T}^{2}}{\mu^{2}}\right)\label{eq:Regge_ff}
\end{equation}
is the so-called Regge form factor. This form has been used in Ref.
\citep{Kutak:2011fu} to propose a non-linear extension of the CCFM
equation. The scale $\mu$ has been introduced to separate unresolved
and resolved emissions in (\ref{eq:BFKL_constr}), i.e. the emissions
with $q_{T}^{2}<\mu^{2}$ and $q_{T}^{2}>\mu^{2}$, and further the
unresolved part was re-summed to obtain the Regge form factor. Note,
that the UGDs undergoing this equation  do not explicitly depend on the scale
$\mu$ and that the new form of the initial condition has to be used
(this is denoted by a tilde sign).

\item BFKL with the kinematic constraint and DGLAP correction (BFKL+CD)

In Ref. \citep{Kwiecinski:1997ee} yet another improvement of (\ref{eq:BFKL})
was proposed. One can make an attempt to account for DGLAP-like behaviour
by including the non-singular part of the gluon splitting function
(the third term below)
\begin{multline}
\mathcal{F}\left(x,k_{T}^{2}\right)=\mathcal{F}_{0}\left(x,k_{T}^{2}\right)\\
+\overline{\alpha}_{s}\left(k_{T}^{2}\right)\int_{x}^{1}\frac{dz}{z}\,\int_{k_{T0}^{2}}^{\infty}dq_{T}^{2}\left[\frac{q_{T}^{2}\mathcal{F}\left(\frac{x}{z},q_{T}^{2}\right)\Theta\left(k_{T}^{2}-zq_{T}^{2}\right)-k_{T}^{2}\mathcal{F}\left(\frac{x}{z},q_{T}^{2}\right)}{\left|q_{T}^{2}-k_{T}^{2}\right|}+\frac{k_{T}^{2}\mathcal{F}\left(\frac{x}{z},k_{T}^{2}\right)}{\sqrt{4q_{T}^{4}+k_{T}^{4}}}\right]\\
+\overline{\alpha}_{s}\left(k_{T}^{2}\right)\int_{x}^{1}\frac{dz}{z}\,\left(\frac{z}{2N_{c}}P_{gg}\left(z\right)-1\right)\int_{k_{T0}^{2}}^{k_{T}^{2}}dq_{T}^{2}\mathcal{F}\left(\frac{x}{z},q_{T}^{2}\right),\label{eq:BFKL+CD}
\end{multline}
where $P_{gg}\left(z\right)$ is the standard gluon splitting function.
This correction, similar to the kinematic constraint, accounts for
certain sub-leading corrections to the BFKL equation.

\item BFKL with DGLAP correction alone\label{enu:lastmodel}

\newcounter{rememblist}\setcounter{rememblist}{\theenumi} This variant
is used to test the significance of the DGLAP term alone.

\end{enumerate}
The above UGDs do not involve any hard scale dependence. For observables
involving high-$p_{T}$ jets a presence of large scale $\mu^{2}\sim p_{T}^{2}$
in perturbative calculations would involve additional logarithms of
the type $\log\left(\mu^{2}/k_{T}^{2}\right)$ which can spoil the
procedure. Therefore a re-summation of those logs is desired and it
accounts in hard scale dependence for UGDs, c.f. Eq. (\ref{eq:HEN_fact_2}).
The approach which incorporates both $x$, $k_{T}^{2}$ and $\mu^{2}$
dependence in UGDs is provided for example by the CCFM evolution equation 
(the code available for a practical use is described for example in \citep{Hautmann:2014uua}).
Another approach, so called KMR (Kimber-Martin-Ryskin) procedure \citep{Kimber:1999xc,Kimber:2001sc},
takes ordinary PDFs and injects $k_{T}$ dependence via the Sudakov
form factor taking care of matching to the BFKL evolution at small
$x$. A serious advantage of this procedure is that one can use well
known PDF sets, fitted to large data sets. Yet another approach was
used in \citep{vanHameren:2014ala} in therms of so-called ``Sudakov
resummation model''. This procedure reverts, in a sense, the logic
used in the KMR and uses the Sudakov form factor to inject the hard
scale dependence instead of $k_{T}$. The procedure is parton-shower-like,
i.e. it is applied after the MC events are generated and the cross
section is known, and is unitary (i.e. the procedure does not change the
total cross section). The advantage is that one may use it on the
top of UGDs involving nonlinear effects. 
The basic idea behind the model is that it assigns the Sudakov probability $P$ for events with given $k_T$ and a hard scale $\mu\sim p_T$. Then, the probability of surviving is $1-P$. For events with small $k_T$ and large $\mu$ the emission probability is $P\sim 1$ and the unitarity of the procedure transfers such events to the region $k_T\sim p_T$.
There is one more approach
proposed in Ref. \citep{Kutak:2014wga}, similar to the one just described,
where analogous procedure is applied at the level of UGDs by fixing
its integral over $k_{T}$ (it has an advantage of being independent on any software and one may produce grids for a practical usage). 
In summary, we may consider the following modifications of UGDs \ref{enu:pure-BFKL-equation}-\ref{enu:lastmodel}:

\begin{enumerate}

\setcounter{enumi}{\therememblist}

\item BFKL with the Sudakov (BFKL+S)

\item BFKL with the kinematic constraint and the Sudakov (BFKL+CS)

\item BFKL with DGLAP correction and the Sudakov (BFKL+DS)

\item BFKL with the kinematic constraint in re-summed form and the Sudakov
(BFKL+CRS)

\item BFKL with the kinematic constraint, DGLAP correction and the Sudakov
(BFKL+CDS)\label{enu:BFKLCDS}

\end{enumerate}

\noindent 
Unfortunately, as far as fitting of UGDs is considered, the above Sudakov-based models are not suitable. This is because they require the knowledge of an integral (whether it is a cross section or integrated gluon, c.f. \citep{vanHameren:2014ala} vs \citep{Kutak:2014wga}) which is unknown at the stage of fitting. In principle, one could try to use the method of successive approximations with the Sudakov model of Ref. \citep{vanHameren:2014ala}. We shall report on our attempts in Section~\ref{sec:Results}.
There is one more comment in order here. The Sudakov resummation
model is very sensitive to the region $k_{T}\lesssim1\,\mathrm{GeV}$
which is not well described by the practical implementations of the
equations \ref{enu:pure-BFKL-equation}-\ref{enu:lastmodel} as they
use certain low-$k_{T}$ cut, $k_{T\,0}$. For $k_{T}<k_{T\,0}$ the
UGD is typically modelled or extrapolated by a constant value.

Let us now discuss the models for the initial condition $\mathcal{F}_{0}$.
In this paper we have tested the following models (in the brackets we give
the aliases used below to identify the model):\renewcommand{\theenumi}{\Alph{enumi}}
\begin{subequations}
\label{eq:param}
\begin{enumerate}
\item exponential model (EXP)\label{enu:Models_ini}
\begin{equation}
\mathcal{F}_{0}\left(x,k_{T}^{2}\right)=N\, e^{-Ak_{T}^{2}}\left(1-x\right)^{a}\left(1-Dx\right)\label{eq:Model_exp}
\end{equation}

\item (negative) power-like model with running $\alpha_{s}$ (POW)
\begin{equation}
\mathcal{F}_{0}\left(x,k_{T}^{2}\right)=\frac{\overline{\alpha}_{s}\left(k_{T}^{2}\right)}{k_{T}^{2}}\, N
\, x^A
\left(1-x\right)^{a}\label{eq:Model_ask}
(1-Dx)
\end{equation}

\item DGLAP-based model (Pgg)\label{enu:Models_fini}
\begin{equation}
\mathcal{F}_{0}\left(x,k_{T}^{2}\right)=\frac{\alpha_{s}\left(k_{T}^{2}\right)}{2\pi k_{T}^{2}}\,
\int_{x}^{1}dz
\, P_{gg}\left(z\right)\hat{\mathcal{G}}_{0}\left(x\right),\label{eq:Model_Pgg_1}
\end{equation}
where 
\begin{equation}
\hat{\mathcal{G}}_{0}\left(x\right)=N
\, x^A
\left(1-x\right)^{a}\left(1-Dx\right)\label{eq:Model_Pgg_2}
\end{equation}
is a model for an integrated gluon density.
\end{enumerate}
\end{subequations}
The parameters $N$, $A$, $a$, $D$ are, in general, free parameters
and need to be fitted.

We see that in principle there are quite a few variants to be fitted.
Though not all of the combinations make sense, we are still left with
several scenarios to be tested.

\section{Fitting procedure}

\label{sec:Procedure}

We have used two data samples measured by CMS detector \citep{Chatrchyan2012}
for inclusive forward-central dijet production at CM energy $\sqrt{s}=7\,\mathrm{TeV}$.
The central jet is defined to lie within the pseudo-rapidity interval
$\left|\eta_{c}\right|<2.8$ while the forward has to lie within $4.9>\left|\eta_{f}\right|>3.2$.
Both jets are high-$p_{T}$ jets with $p_{T}>35\,\mathrm{GeV}$. The
jets were reconstructed using anti-$k_{T}$ algorithm with radius
$R=0.5$. The data samples consist in jet $p_{T}$ spectra for forward
and for central jets, $d\sigma_{S}/dp_{T}\Delta\eta_{S}$ with $S=f,c$.
There are in total 12 data bins for both forward and central jets. 

We have applied the following fitting procedure. For each existing
experimental data bin $B$ we produce a 2-dimensional normalized histogram
$\mathcal{H}^{B}$ with bins in $x$ and $k_{T}$, such that the cross
section can be calculated as
\begin{equation}
\sigma^{B}=\sum_{i,j}\mathcal{H}_{ij}^{B}\mathcal{F}\left(x\left(i\right),k_{T}\left(j\right)\right),\label{eq:Fit1}
\end{equation}
where $i,j$ enumerate the bins in $\left(x,k_{T}\right)$. To make
the histograms $\mathcal{H}^{B}$ we\renewcommand{\theenumi}{\arabic{enumi}}
\begin{enumerate}
\item generate Monte Carlo events for the process under consideration with
$\mathcal{F}=\mathcal{F}^{*}$, where $\mathcal{F}^{*}$ is a relatively
``broad'' trial UGD (evolving according to one of the scenarios
\ref{enu:pure-BFKL-equation}-\ref{enu:BFKLCDS}),
\item make histograms $\mathfrak{h}^{B}$ in $\left(x,k_{T}\right)$ of
contributions to each data bin $B$,
\item divide by $\mathcal{F}^{*}\left(x,k_{T}\right)$, i.e. $\mathcal{H}_{ij}^{B}=\mathfrak{h}_{ij}^{B}/\mathcal{F}^{*}\left(x\left(i\right),k_{T}\left(j\right)\right)$.
\end{enumerate}
Hence, in principle $\mathcal{H}^{B}$ are independent of $\mathcal{F}^{*}$
used for their generation and are calculated only once. This is advantageous,
as the hard cross section calculation is costly in CPU time. The latter
is calculated using the Monte Carlo C++ program $\mathtt{LxJet}$
\citep{Kotko2013a} implementing (\ref{eq:HEN_fact_2}). The generated
events (weighted or unweighted) are stored in a $\mathtt{ROOT}$ \citep{Brun:1997pa}
file for further processing. For the UGD evolution according to scenarios
\ref{enu:pure-BFKL-equation}-\ref{enu:lastmodel}
we solve the corresponding integral equations by a straightforward numerical iteration
over a grid over $x$ and $k_T$.

In order to make the fitting feasible, we need a fast routine to calculate $\mathcal{F}$
used in (\ref{eq:Fit1}) for
the cross section calculation. However, since 
our numerical procedure is too slow for that,
we prepare grids over which we can interpolate the fitting parameters.
Each such grid corresponds to a particular
parametrization model and arguments range.
Out of four parameters ($N$, $A$, $a$, $D$) of the initial conditions, we fix $D=0$ (see Sec. \ref{sec:Results}).
Moreover, we note that the solution for $\mathcal{F}$ is linear in $N$. Thus the actual grids are in $A$ and $a$.

\section{Results}

\label{sec:Results}

\begin{figure}
\begin{center}
\includegraphics[clip,width=0.48\textwidth]{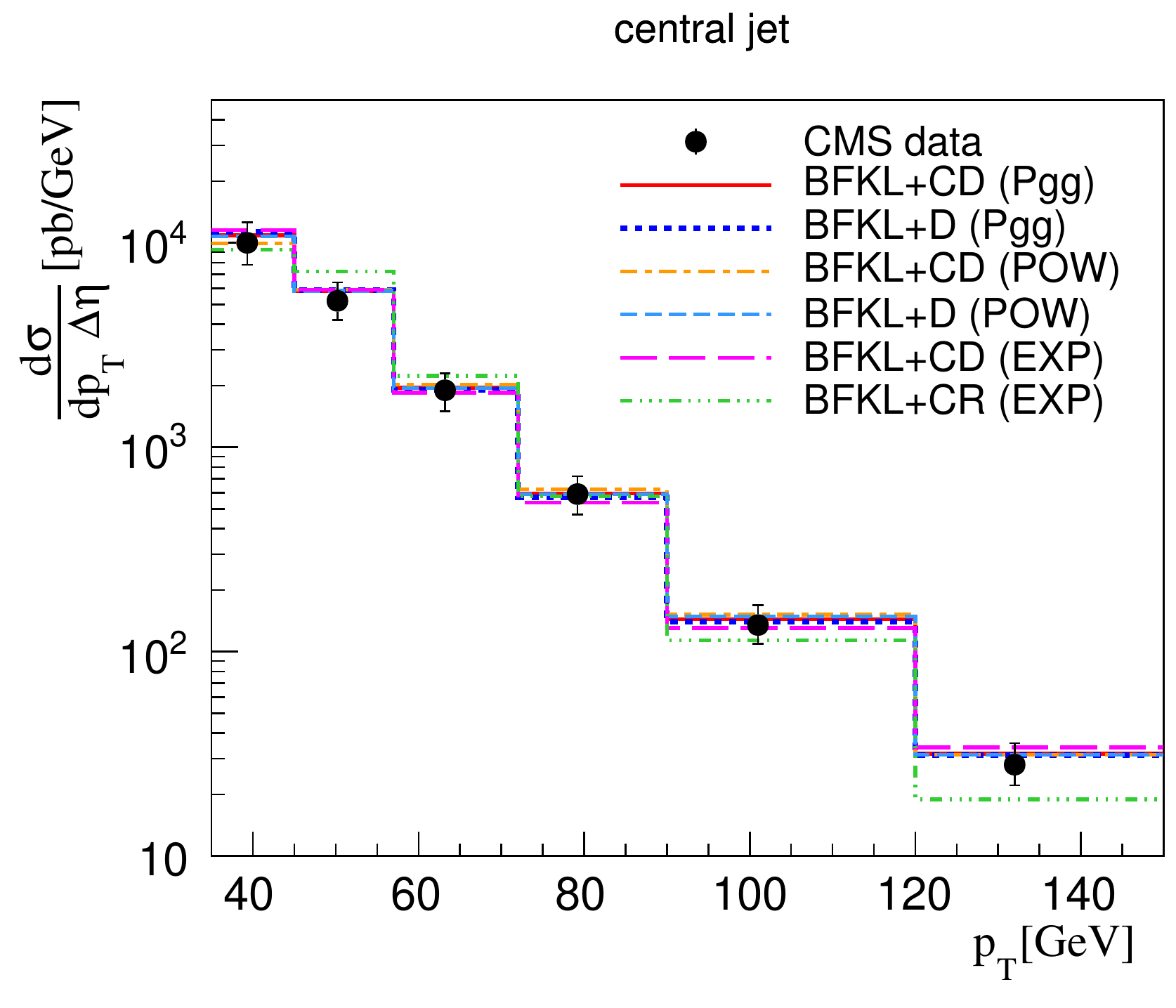}
\hfill
\includegraphics[clip,width=0.48\textwidth]{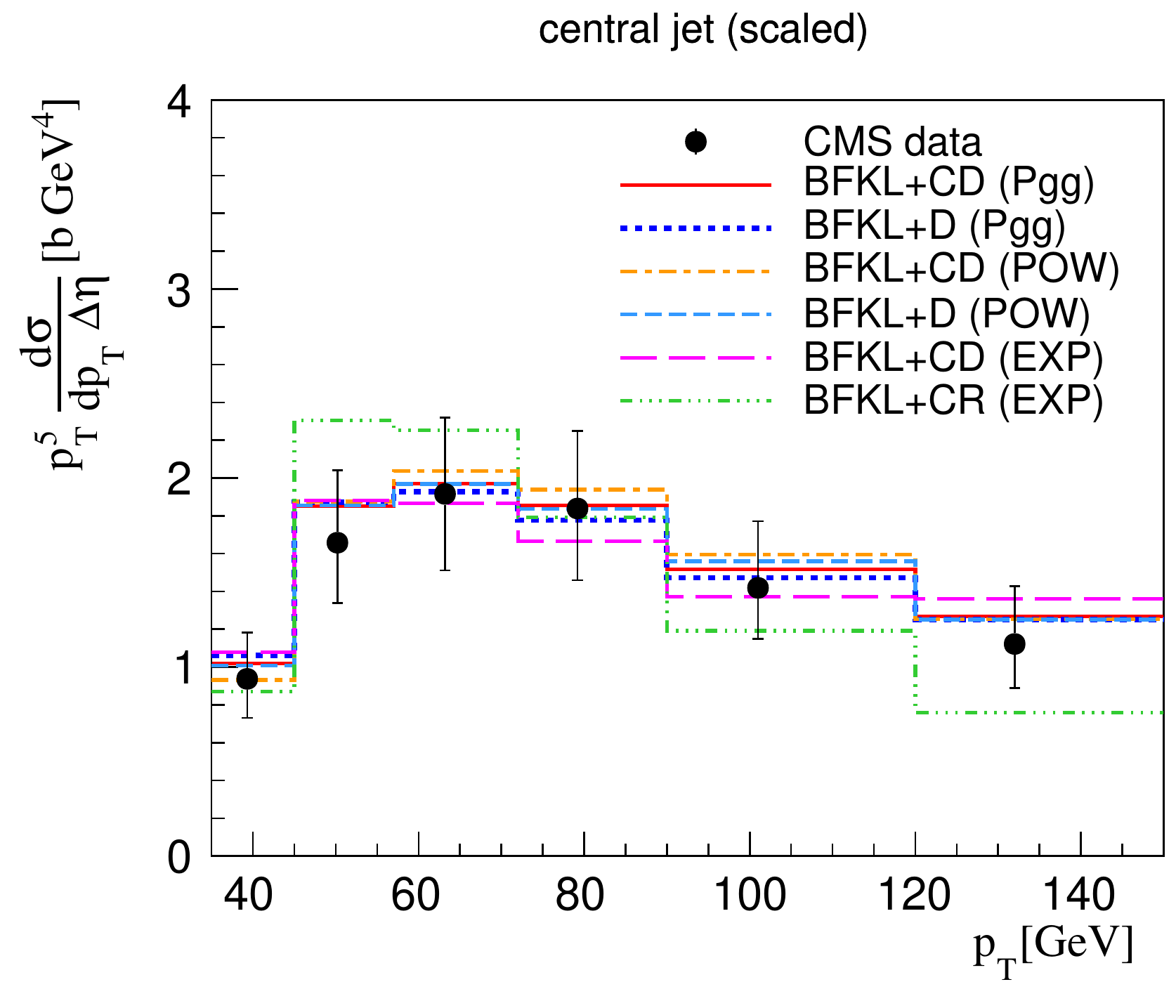}
\par\end{center}

\caption{The $p_{T}$ spectra of the central jet calculated using the best
fits for individual models versus the CMS data. For the bottom plot
the cross sections have been scaled by $p_{T}^{5}$ to better see
the differences between the models.\label{fig:central}}

\end{figure}

\begin{figure}
\begin{center}
\includegraphics[clip,width=0.48\textwidth]{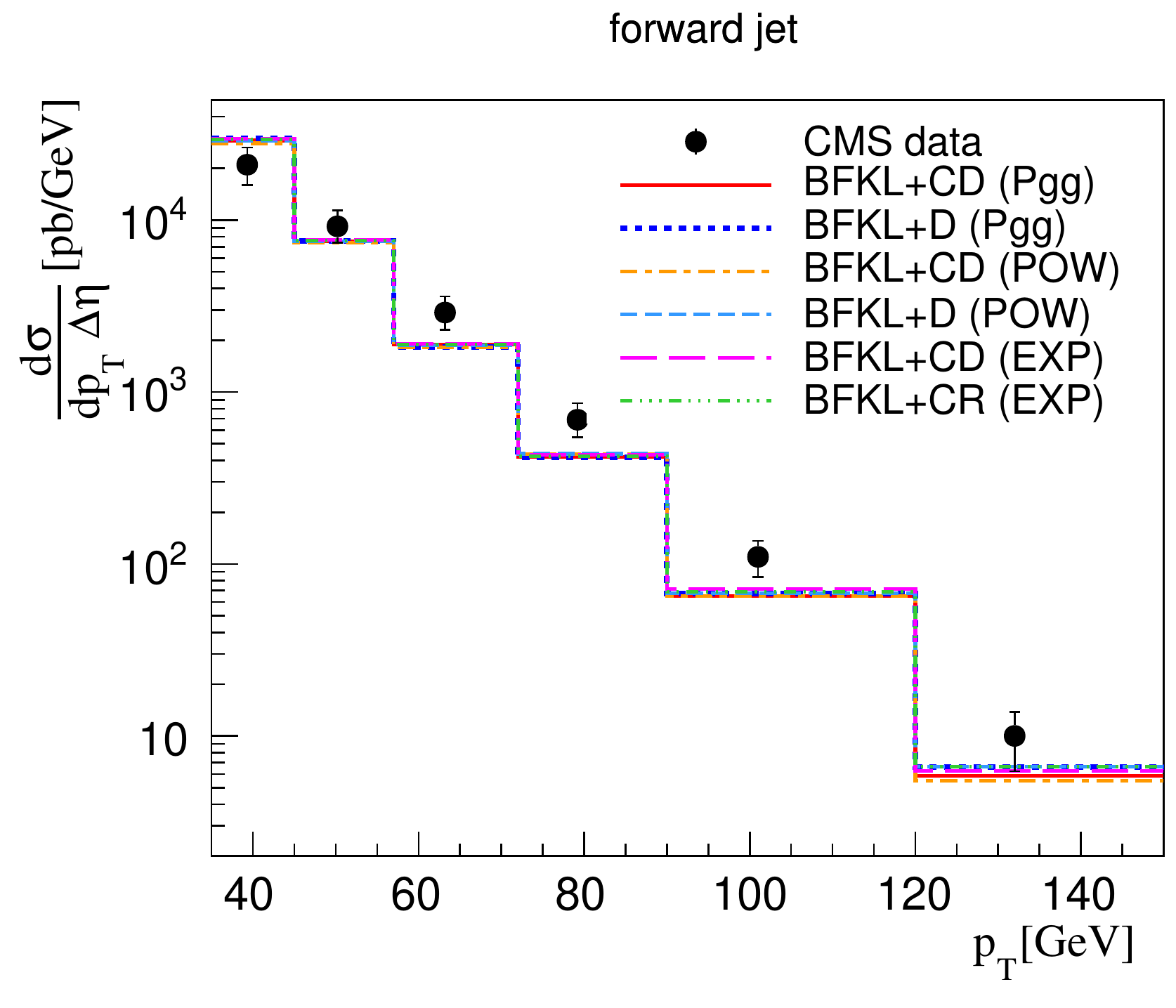}
\hfill
\includegraphics[width=0.48\textwidth]{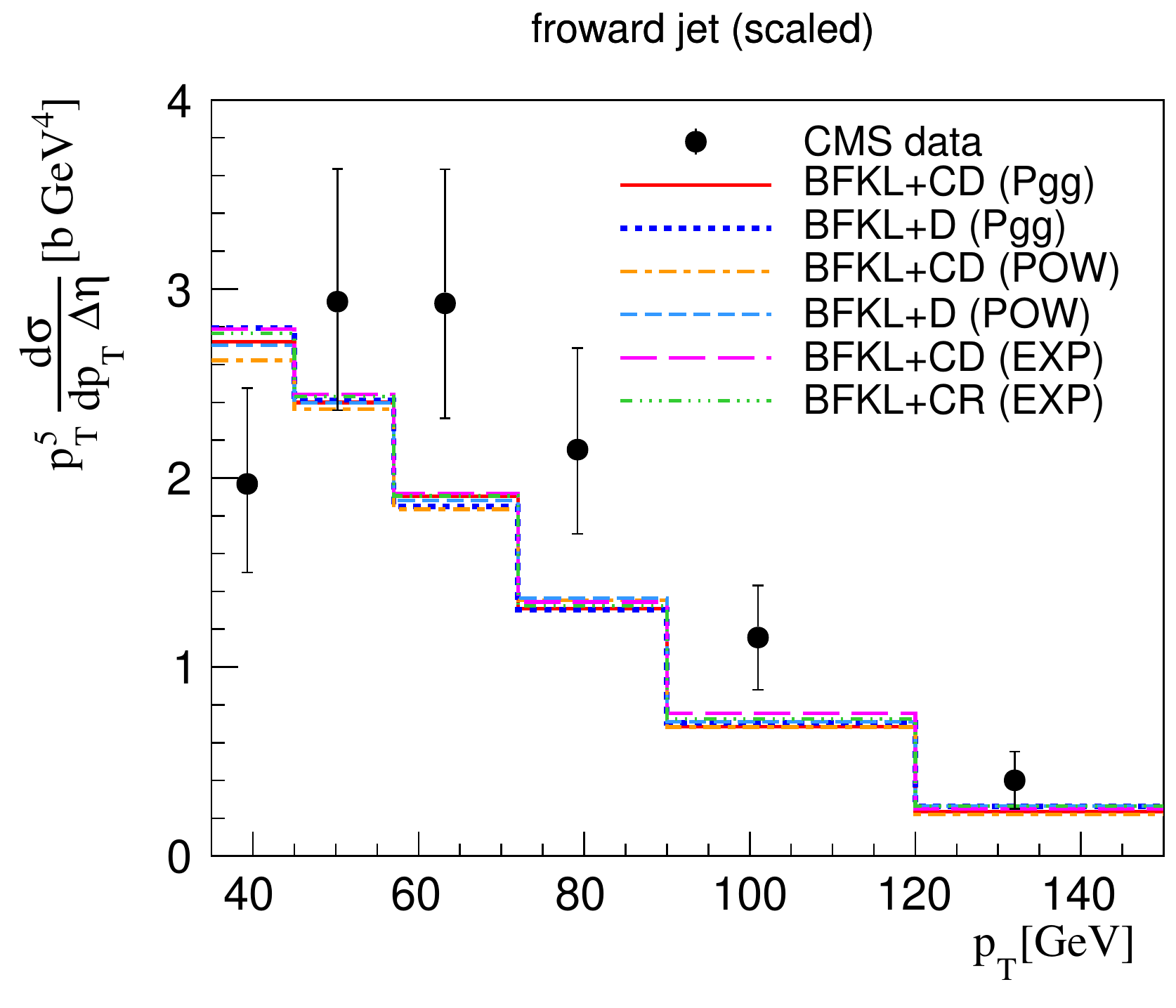}
\par\end{center}

\caption{The $p_{T}$ spectra of the forward jet calculated using the best
fits for individual models versus the CMS data. For the bottom plot
the cross sections have been scaled by $p_{T}^{5}$ to better see
the differences between the models.\label{fig:forward}}
\end{figure}

We have applied the procedure described in the preceding section to
most of the models \ref{enu:pure-BFKL-equation}-\ref{enu:BFKLCDS}
and initial conditions \ref{enu:Models_ini}-\ref{enu:Models_fini}.
The best values of \chisq ($\chi^2$ per data point) are listed in Table \ref{tab:chi2}
for models \ref{enu:pure-BFKL-equation}-\ref{enu:lastmodel}. 
Note, that some of the scenarios were unable to describe the
data, in particular the pure BFKL and BFKL with the kinematic constraint only. 
Evidently, the DGLAP correction is essential.  The fitted values of the parameters of the initial conditions, $N$, $A$,
$a$, for scenarios with $\chisq<2$ are collected in Table \ref{tab:params}.
The fits are
presented in Figs. \ref{fig:central}-\ref{fig:forward}.
For a better comparison we also plot the cross-sections scaled by $p_T^5$.
We observe that all the models with the DGLAP correction give excellent description of the central-jet data,
while the $p_T$ spectrum of forward jets
is reasonably reproduced though less accurately.
We also note that the models with lowest $\chi^2$ result in very similar predictions for the $p_T$ spectra.

\begin{figure}
\begin{center}
\includegraphics[width=0.48\textwidth]{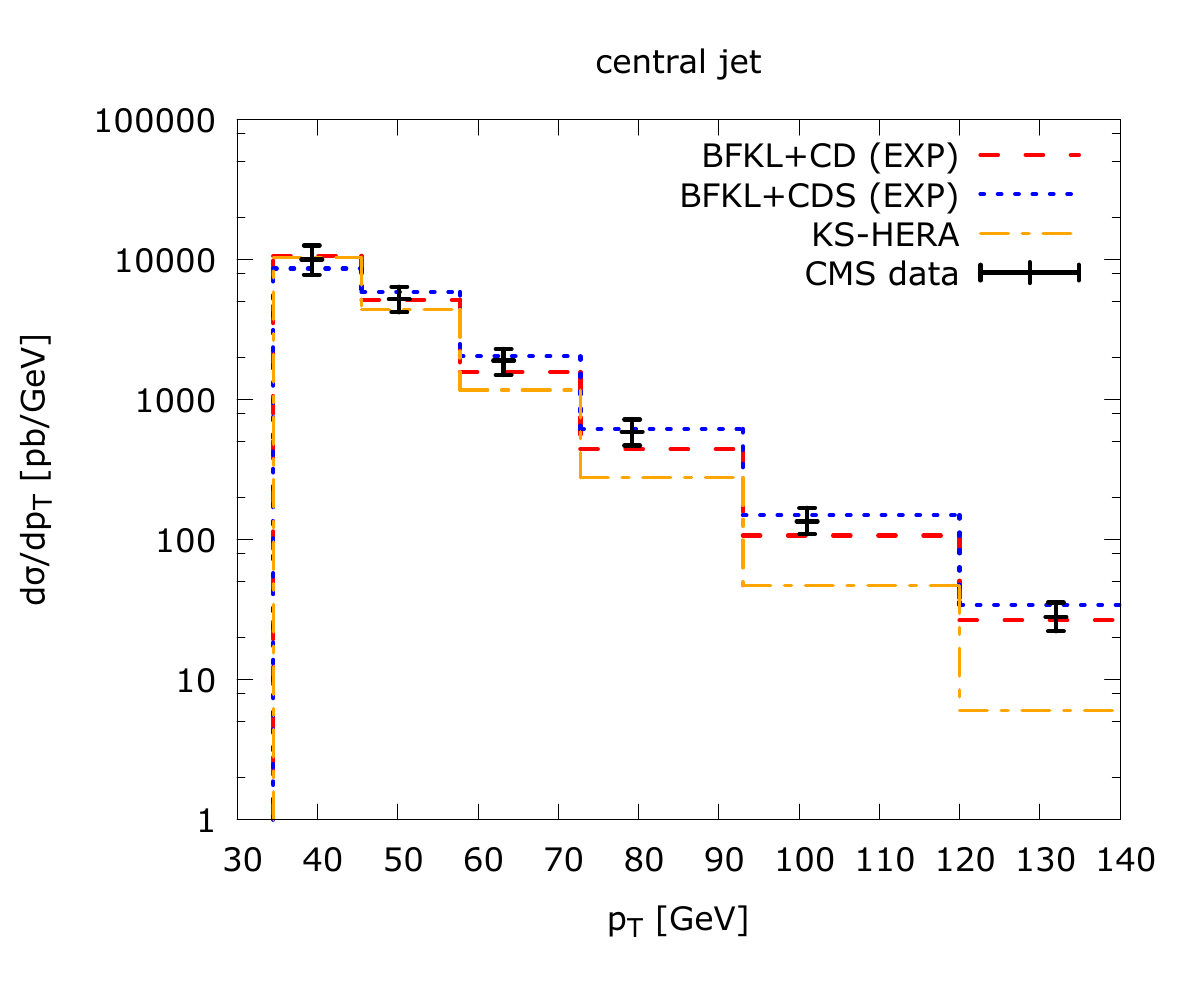}
\hfill
\includegraphics[width=0.48\textwidth]{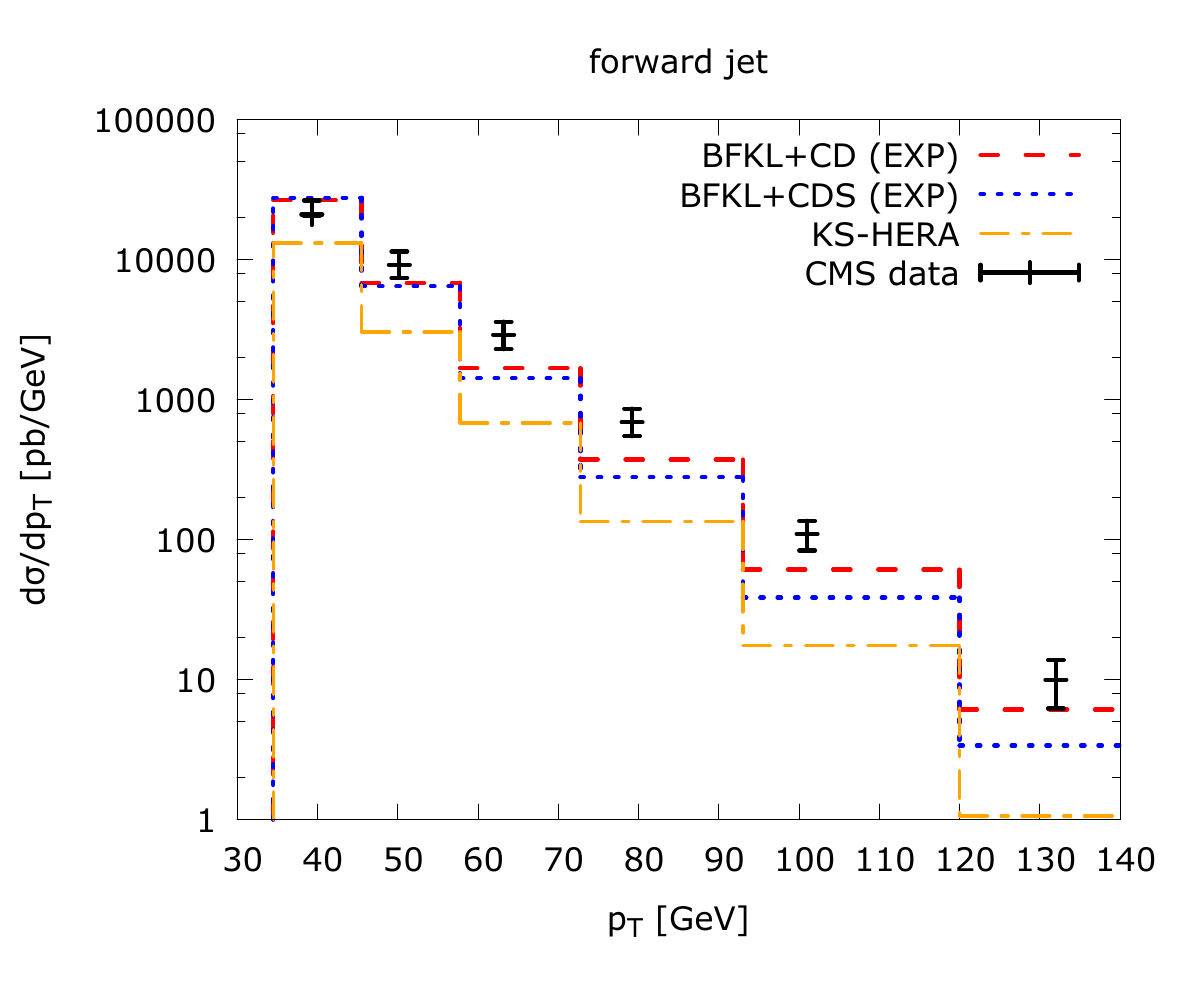}
\par\end{center}

\caption{An effect of the Sudakov resummation model (BFKL+CDS) when applied to one of our fits for the model BFKL+CD with exponential initial condition. For comparison we plot also the spectra obtained from the unintegrated gluon density with more involved evolution and fitted to HERA data (KS-HERA), see the main text for more details. 
\label{fig:sudeffect}}
\end{figure}

Our attempts to fit the scenarios with the Sudakov resummation can be summarized as follows. First, we observe that the model has a small overall effect on the $p_T$ spectra, although it slightly shifts the theory points away from the data points. We illustrate this in Fig.~\ref{fig:sudeffect}, where we applied the Sudakov model on the top of the events obtained with one of the fits. When we now try to refit the $\mathcal{F}_0$ parameters, we change the total cross section (used already to apply the resummation) and the fit fails.
Although we observe that the successive iterations improve the fit, the procedure turns out to be insufficient to make a reliable fit with the Sudakov resummation.

\begin{table}
\begin{center}
\doublespacing
\begin{tabular}{|c|c|c|c|c|c|}
\hline 
$\mathcal{F}_{0}$ & BFKL & BFKL+C & BFKL+D & BFKL+CD & BFKL+CR\tabularnewline
\hline 
\hline 
EXP & 2.4 & 2.2 & 1.24 & 1.11 & 1.52\tabularnewline
\hline 
POW & 2.3 & 1.9 & 1.02 & 1.12 & \tabularnewline
\hline 
Pgg & -- & -- & 1.13 & 1.11 & \tabularnewline
\hline 
\end{tabular}
\par\end{center}

\caption{The values of \chisq for fits of unintegrated gluon density
evolving according to various models described in Section \ref{sec:evolution_eqs}.
The first column lists the initial condition ansatz, see also Section~\ref{sec:evolution_eqs} for details.\label{tab:chi2}}

\end{table}

\begin{table}
\begin{center}
\doublespacing
\begin{tabular}{|c|c|c|c|}
\hline 
model & $N$ & $A$ & $a$\tabularnewline
\hline 
\hline 
BFKL+CR (EXP) & $0.095$ & $0.012$ & $0^*$\tabularnewline
\hline 
BFKL+D (EXP) & $0.37$ & $0.18$ & $0.5^*$\tabularnewline
\hline 
BFKL+CD (EXP) & $0.68$ & $0.14$ & $2.5^*$\tabularnewline
\hline 
BFKL+C (POW) & $320$ & $1.4$ & $61.0$\tabularnewline
\hline 
BFKL+D (POW) & $12.7$ & $0.5^*$ & $5.7$\tabularnewline
\hline 
BFKL+CD (POW) & $562$ & $0.96$ & $35.7$\tabularnewline
\hline 
BFKL+D (Pgg) & $106$ & $1.2$ & $2.5$\tabularnewline
\hline 
BFKL+CD (Pgg) & $628$ & $2.9$ & $5.7$\tabularnewline
\hline 
\end{tabular}
\par\end{center}

\caption{The values of initial condition \ref{enu:Models_ini}-\ref{enu:Models_fini}
parameters obtained from the fits to the CMS data. We list only the scenarios
with $\chisq < 2$. The values denoted by a star were fixed --- see the main text for details.\label{tab:params}}
\end{table}

A few comments are in order.
The considered jet data are not sufficient to precisely determine all the parameters
($N$, $A$, $a$, $D$)
of the initial parametrizations (\ref{eq:param}).
Thus, first we
neglect the $(1-Dx)$ factor, i.e. we take $D=0$.
We have checked that we get no improvement when $D$ is a free parameter.
Next, in some cases the fits are not sensitive enough to uniquely determine the three remaining
free parameters. In these cases we fix $A$ or $a$ at some plausible value
(these are marked with a star in Table \ref{tab:params}).
Actually, besides the initial condition
parameters $N$, $A$, $a$, $D$ we have also the boundary values
of kinematic parameters $x_{A}$, $k_{T}$ (c.f. (\ref{eq:HEN_fact_2})),
which -- to certain extent -- are free parameters as well. We set
them as follows. First, in order to be in an accordance with the assumptions
leading to (\ref{eq:HEN_fact_2}) we imply the cut $x_{A}<x_{B}$.
Next, for all scenarios we set $x_{A\,\mathrm{min}}=0.0001$. For
the model with the DGLAP correction we set $x_{A\,\mathrm{max}}=1.0$
while for the others we set $x_{A\,\mathrm{max}}=0.4$. Further we
use $k_{T\,\mathrm{min}}=1\,\mathrm{GeV}$ for DGLAP models and $k_{T\,\mathrm{min}}=0.1\,\mathrm{GeV}$
for the others. Finally, we use $k_{T\,\mathrm{max}}=100\,\mathrm{GeV}$
for exponential initial condition and $k_{T\,\mathrm{max}}=400\,\mathrm{GeV}$
for the others. The last comment concerns the hard scale choice: in
all fits we have used the average $p_{T}$ of the jets.

The influence of the Sudakov resummation model is illustrated in Fig.~\ref{fig:sudeffect}. Here, we have chosen the best fits to illustrate
the effect. We see, that the jet spectra are rather weakly affected
by the resummation, although the forward jet spectrum becomes steeper
than the data.

\begin{figure}
\begin{center}
\includegraphics[width=0.9\textwidth]{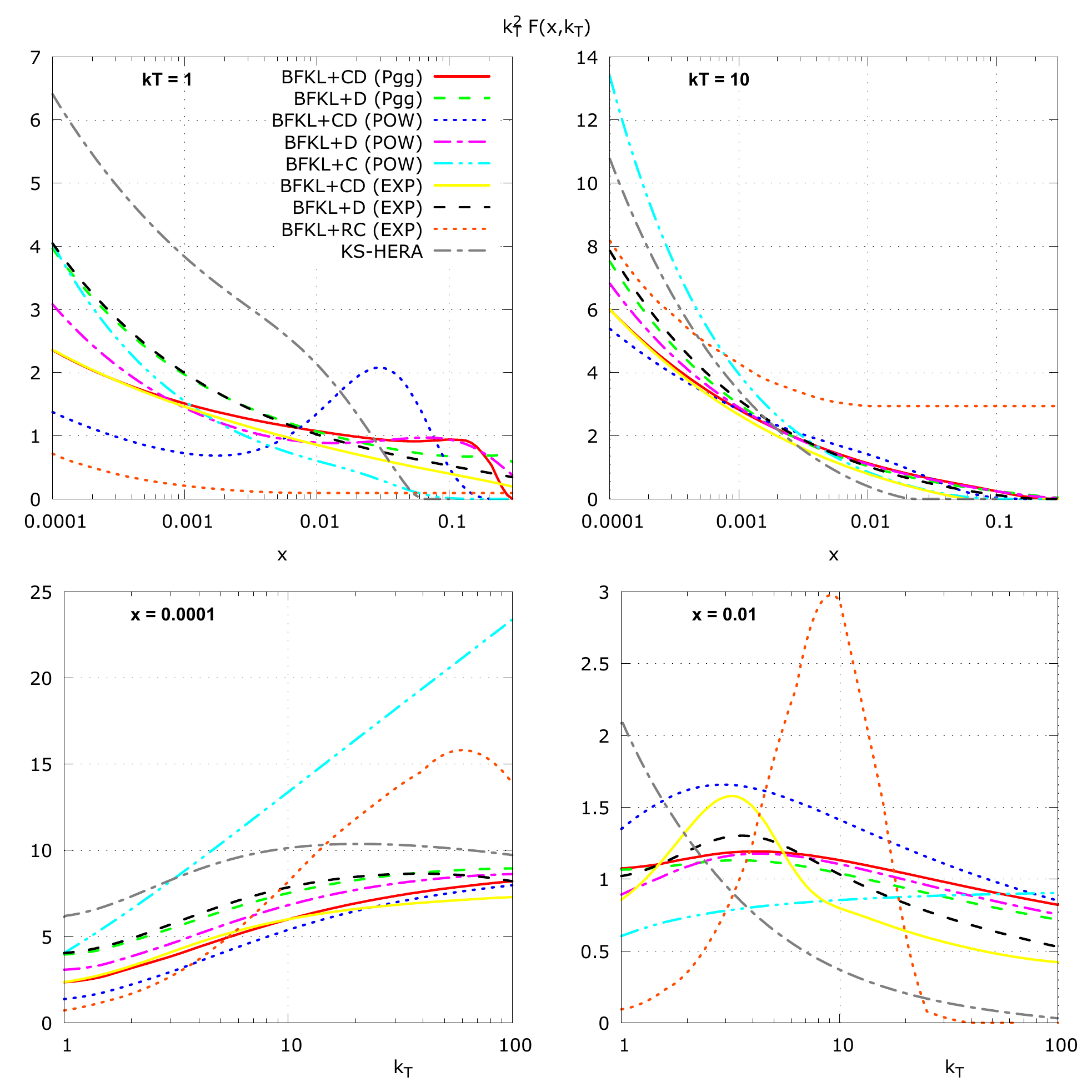}
\end{center}
\caption{Unintegrated gluon distributions evolving due to the models \ref{enu:pure-BFKL-equation}-\ref{enu:lastmodel} with the  initial conditions \ref{enu:Models_ini}-\ref{enu:Models_fini} obtained from the fits to the LHC data 
as a function of $x$ (top) and $k_T$ (bottom). The UGDs are multiplied by $k_T^2$ to better illustrate the differences between the models. The most differing UGDs are those without the DGLAP correction and with significantly higher $\chisq > 1.5$ (BFKL+C and BFKL+RC). 
\label{fig:gluons}}
\end{figure}

The obtained UGDs are plotted in one-dimensional plots in Fig.~\ref{fig:gluons} as a function of $x$ and $k_T$. Note, that in order to better reflect the difference between UGDs we plot $k_T^2\, \mathcal{F}(x,k_T)$.
We show results of all the models of Table \ref{tab:params}, hence also those with rather high $\chi^2$ value (see Table \ref{tab:chi2}).
All the UGDs with the DGLAP contribution are comparable, which shows that the evolution scenario is more important than a particular shape of the initial parametrization.
On the other hand, the differences between UGDs are more pronounced than those in the $p_T$ spectra,
which means that the currently available data are not sufficient to discriminate among the models.
The two most differing UGDs correspond to the BFKL+C (POW) and BFKL+CR (EXP) models
which however have significantly higher $\chisq$ (above 1.5).

We compare the new LHC-based UGDs with the one evolving according to a complicated evolution of \cite{Kwiecinski:1997ee,Kutak:2004ym} and  fitted to HERA data \cite{Kutak:2012rf} (we abbreviate it as 'KS-HERA' on the figure). This evolution equation contains the kinematic constraint, full DGLAP correction (including quarks via coupled equations) and a nonlinear term motivated by the Balitsky-Kovchegov equation.
The $p_T$ spectra resulting from this gluon density are presented in Fig.~\ref{fig:sudeffect}.

\section{Azimuthal decorrelations}

\label{sec:decorrelations}

In order to apply the fits in practice we have calculated another
observable for central-forward dijet production, namely, the differential
cross sections in azimuthal angle $\Delta\phi$ between the two jets.
At leading order the two jets are produced exactly back-to-back and
the distribution is the Dirac delta at $\Delta\phi=\pi$.
However, due to QCD emissions of additional partons (either forming
additional jets or being soft particles with small $p_{T}$) the two
jets are decorrelated. On theory side these decorrelations are well
described by QCD-based parton shower algorithms. However, within the
HEF there is a natural decorrelation mechanism built-in. Namely, due
to the internal transverse momentum $k_{T}$ of a gluon the dijet
system with transverse momenta $\vec{p}_{T1}$, $\vec{p}_{T2}$ is
unbalanced by the amount $\left|\vec{p}_{T1}+\vec{p}_{T1}\right|=\left|\vec{k}_{T}\right|=k_{T}$.
One can think of $k_{T}$ as a cumulative transverse momentum of many
gluon emissions. In general, these emissions can be small-$p_{T}$
and large-$p_{T}$ emissions as well. The large-$p_{T}$ emissions
may in general contribute a jet, thus we consider an inclusive dijet
observables. 

\begin{figure}
\begin{center}
\includegraphics[width=0.5\textwidth]{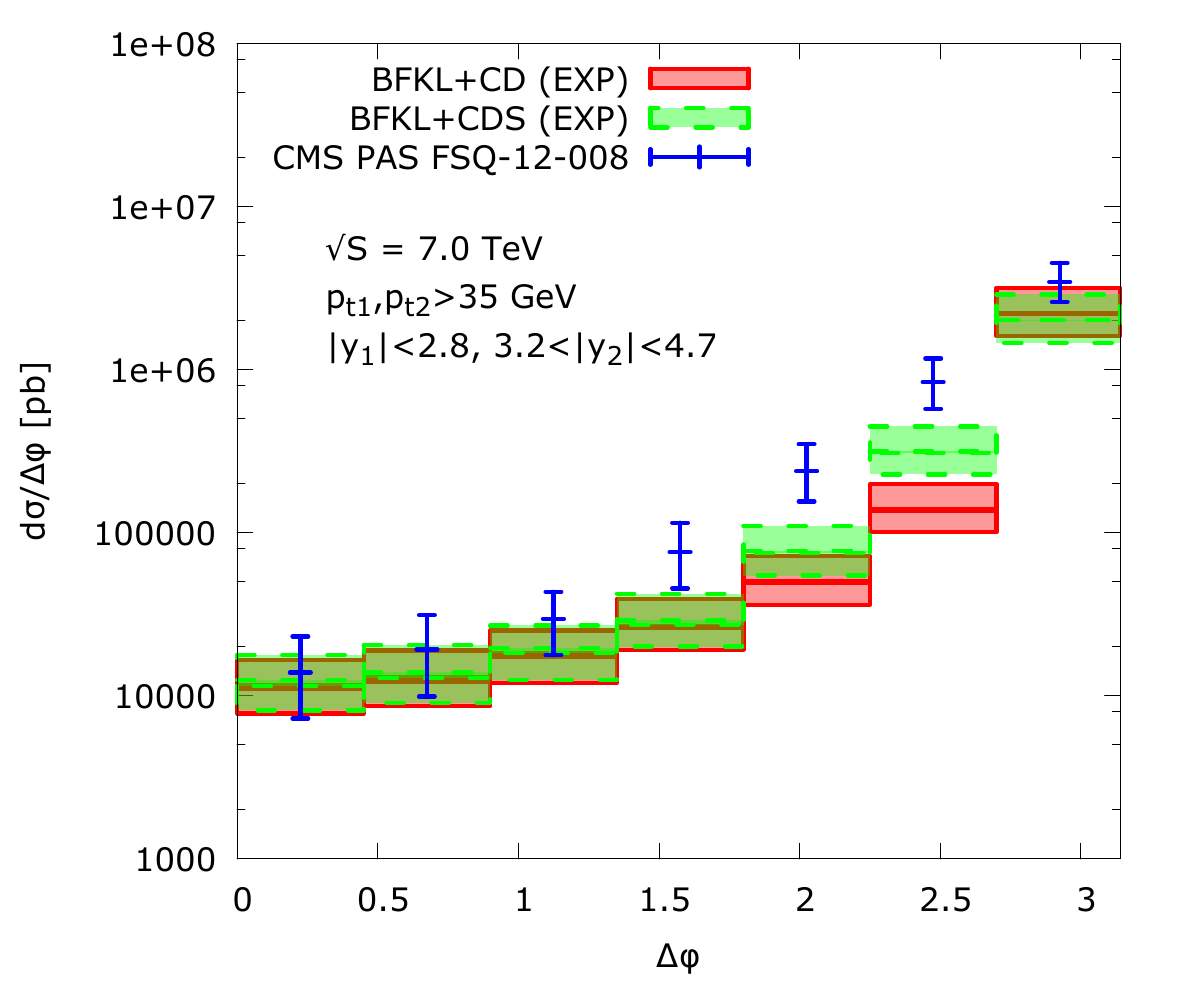}\includegraphics[width=0.5\textwidth]{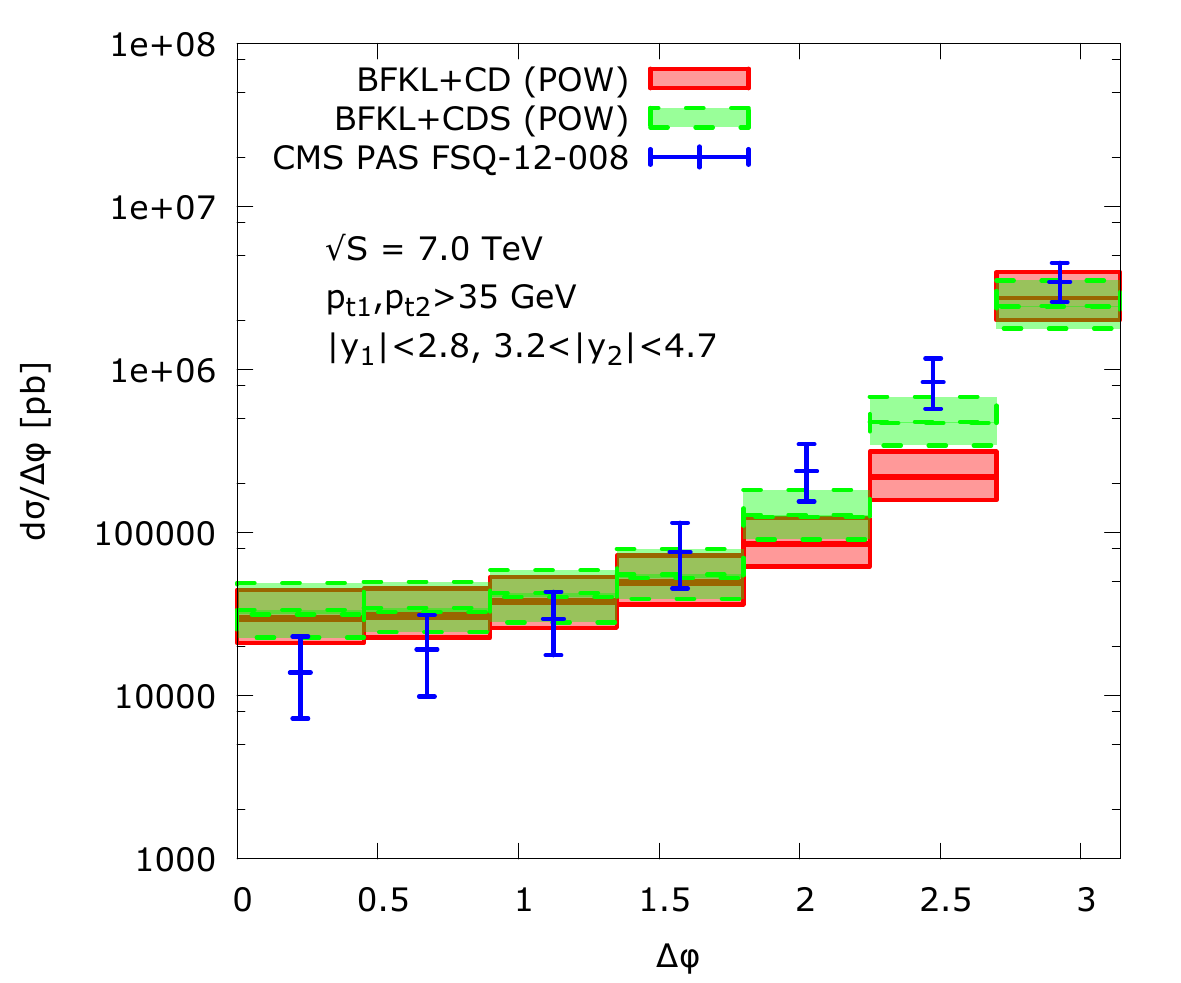}
\par\end{center}

\begin{center}
\includegraphics[width=0.5\textwidth]{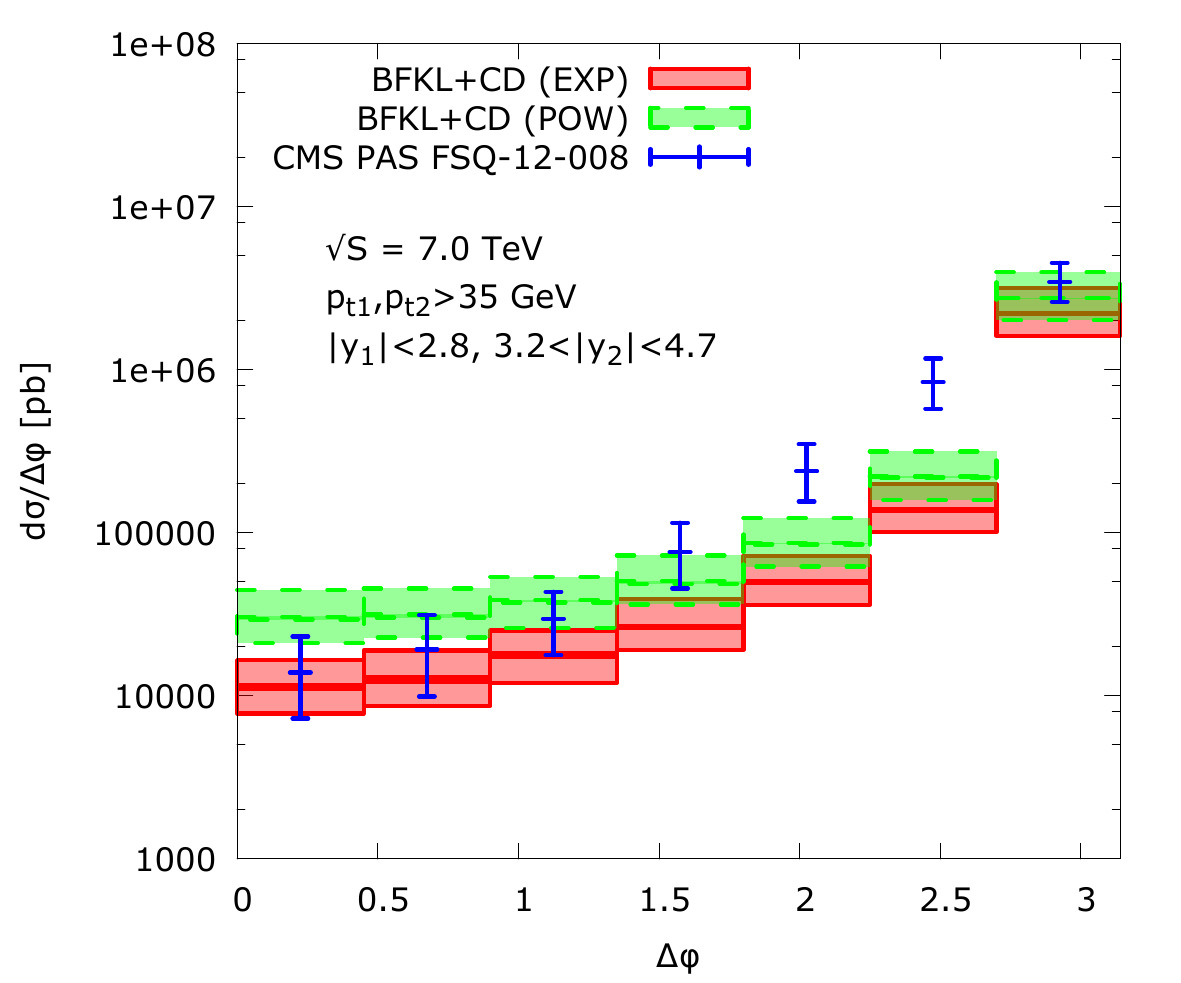}\includegraphics[width=0.5\textwidth]{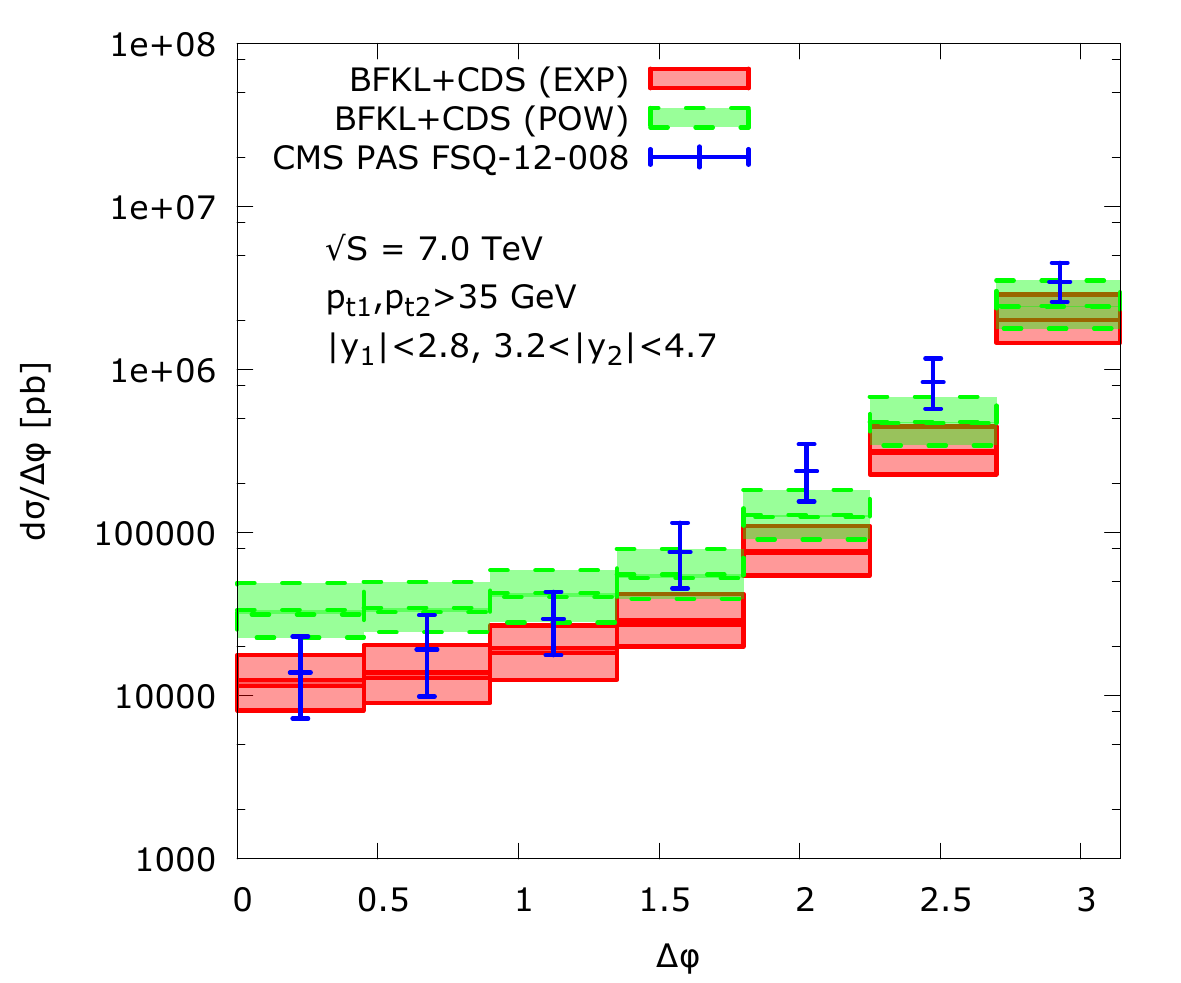}
\par\end{center}

\caption{The results for the azimuthal decorrelations for inclusive forward-central
dijet production using our best fits. When the Sudakov resummation
model is applied to the generated events we get a better description
of the CMS data.\label{fig:decorr}}
\end{figure}

Using the new fits and the $\mathtt{LxJet}$ program we have calculated
the azimuthal decorrelations for the kinematics described in the beginning
of Section \ref{sec:Procedure}. The results are presented in Fig.
\ref{fig:decorr}. The bands represent uncertainty  that comes
from the scale variation by a factor of two. We compare our calculation
with the preliminary CMS data \citep{CMS:2014oma}%
\footnote{We note that the total cross section obtained from \citep{CMS:2014oma}
does not agree with \citep{Chatrchyan2012}. The ratio of the
two is approx.~1.8.
If this is a normalization difference only, our predictions should be shifted up by this factor.
}.

\section{Discussion}

\label{sec:Summary}

In the present paper we went through a thorough study of various small-$x$
evolution equations analyzing an impact of various effects on jet observables.
The effects we mean here, are certain sub-leading corrections to the
BFKL equation, such as the kinematic constraint or DGLAP corrections.
Our study was based on fitting these evolution scenarios to two samples
of LHC data for high-$p_{T}$ spectra for dijet production. These
samples consist of separate spectra for the central rapidity and forward
rapidity jets. 

Our findings can be summarized as follows. First observation is that
both forward jet and central jet spectra can be simultaneously and
reasonably described by the High Energy Factorization approach and
BFKL-like evolution. We obtain the best quality fits for BFKL with
DGLAP correction and kinematic constraint, with the DGLAP correction being the most important additional ingredient. 
This matches the fact that the data under consideration can be nicely described by the collinear factorization with a parton shower \citep{Chatrchyan2012,CMS:2014oma}. Whereas in the High Energy Factorization the parton shower is -- to some extent -- simulated by the transverse momentum dependent gluon distribution with the DGLAP correction.
For all evolution models
we get very good fits to the central jet spectrum, while most of the
models have problems with 
precise reproduction of the shape of the forward jet spectrum.
Several models properly describe the dijet data despite some differences in the resulting UGDs.
Measurements of some other observables or more differential dijet data could help to discriminate among the models.

Using our fits we have calculated azimuthal decorrelations for
the same kinematic domain. This observable was also measured by CMS.
The comparison of our calculation with the data is reasonably good,
especially when using the Sudakov resummation model on the top of
the evolution models. Interestingly, the same resummation procedure
spoils the forward jet $p_{T}$ spectrum.

Our final remark is that although the High Energy Factorization with
improved BFKL evolution equation catches the main physical aspects
of the jet production at small $x$, one  definitely needs higher order
corrections. Such calculations exist for certain small $x$ processes
like Mueller-Navelet jets \citep{Ducloue:2013hia,Ducloue:2014koa}
or inclusive hadron production p+A collisions within CGC formalism
\citep{Chirilli:2012jd,Staato2014}, but not for the high-$p_{T}$ dijet observables
under consideration.

\section*{Acknowledgments}

We thank K.~Kutak and A.~van~Hameren for many fruitful discussions.
The work of P.K. and W.S. has been supported by the
Polish  National Science Center Grant No. \mbox{DEC-2011/03/B/ST2/00220}.
D.T. has been supported by NCBiR Grant No. \mbox{LIDER/02/35/L-2/10/NCBiR/2011}.
P.K. also acknowledges the support of DOE grants No. \mbox{DE-SC-0002145} and \mbox{DE-FG02-93ER40771}.

\bibliographystyle{chetref}
\bibliography{library}

\end{document}